\documentclass[12pt]{article}
\usepackage{latexsym}
\usepackage{cite}
\usepackage{mathrsfs}

\date{}
\newcommand{\eref}[1]{(\ref{#1})}
\newcommand{\rmd}{\mathrm{d}}

\title{\Large \textbf{Anti-self-dual gravity from asymmetric heavenly equation standpoint}}
\author{\large\textbf{M B Sheftel$^{1}$ and D Yaz{\i}c{\i}$^2$}\\
\small $^1$ Department of Physics, Bo\u{g}azi\c{c}i University, Bebek, 34342 Istanbul, Turkey\\
\small $^2$ Department of Physics, Y{\i}ld{\i}z Technical University, Esenler 34220 Istanbul\\ \hspace*{-11.4cm} \small Turkey\\
\hspace*{-2.8cm}\small E-mail: mikhail.sheftel@boun.edu.tr, yazici@yildiz.edu.tr}

\begin{document}
\maketitle

\begin{abstract}
In paper \cite{shma} on the classification of second-order PDEs with four independent variables that possess partner symmetries, asymmetric heavenly equation appears as one of canonical equations admitting partner symmetries. It was shown that all these canonical equations, together with general heavenly equation of Dubrov and Ferapontov \cite{fer}, provide potentials for anti-self-dual Ricci-flat vacuum metrics \cite{pleb,husain,ms}, the asymmetric heavenly equation presenting the only exception so far. Our aim here is to show that the latter equation also governs anti-self-dual vacuum heavenly metric. We present the corresponding basis of null vector fields, null tetrad of coframe 1-forms and the general form of the metric. We obtain a multi-parameter polynomial solution of our equation which yields a family of metrics with the above properties. Riemann curvature 2-forms are also explicitly presented for the cubic solution to
modified heavenly equation in \cite{fer}, which is a particular case of the asymmetric heavenly equation.
\end{abstract}
\begin{flushleft} \small
\hspace*{9mm} PACS numbers: 02.30.Jr, 02.40.Ky, 04.20.Jb\\ \small
\hspace*{9mm} AMS classification scheme numbers: 35Q75, 83C15
\end{flushleft}

\section{Introduction}
\label{sec-intro}

In his pioneering paper \cite{pleb}, Pleba\~nski showed that for anti-self-dual (ASD) metrics complex Einstein equations reduce to a single scalar partial differential equation (PDE) for a potential of the metric, namely, first and second heavenly equations of Pleba\~nski. After imposing reality conditions
on the complex K\"ahler metric, the first heavenly equation becomes either elliptic or hyperbolic complex Monge-Amp\`ere equation governing K\"ahler-Einstein metric with Euclidean or neutral (ultrahyperbolic) signature, respectively.
Later Husain presented another scalar PDE which determined a potential of an ASD Ricci-flat metric \cite{husain}.
In our paper with Malykh \cite{shma}, we classified with respect to point and Legendre transformations all scalar second-order PDEs with four complex variables, which possess partner symmetries (and as a consequence admit a two-dimensional divergence form) and contain only second derivatives of the unknown.
The two heavenly equations of Pleba\~nski and the Husain equation, together with mixed heavenly equation related to the latter one, arise as canonical equations in this classification. The general heavenly equation of Dubrov and Ferapontov \cite{fer} does not appear in our classification because it admits only a three-dimensional divergence form and hence does not possess partner symmetries (in the usual sense) to serve as a tool for its integration. Recently we had obtained a description of ASD gravity in terms of solutions of the latter equation together with its real forms corresponding to Euclidean and neutral signature \cite{ms}.

The only remaining canonical equation in our classification as well as in the classification of \cite{fer}, whose relation to ASD
gravity has not yet been studied, is the asymmetric heavenly equation
\begin{equation}\label{asym}
    u_{14}u_{24} - u_{12}u_{44} + au_{34} + bu_{13} + cu_{11} = 0
\end{equation}
together with modified heavenly equation of Dubrov and Ferapontov \cite{fer}, which is a special case of \eref{asym} when $a=c=0$, $b=1$.
Here and later on $u=u(z^1,z^2,z^3,z^4)$ and $u_{ij}=\partial^2u/\partial z^i\partial z^j$. Bi-Hamiltonian structure of \eref{asym} was recently discovered in \cite{dy}.

The purpose of this study is to discover the relation between the asymmetric heavenly equation and ASD gravity and thus
fill in the gap in the description of ASD gravity in terms of solutions of heavenly equations.

In section \ref{sec-lax}, we present a Lax pair for the asymmetric heavenly equation which is a slight generalization of the Lax pair in \cite{fer}.
A null tetrad for ASD vacuum metric governed by this equation is constructed here by satisfying conditions of the Ashtekar-Jacobson-Smolin-Mason-Newman theorem \cite{ajs,mn,MasWood}.

In section \ref{sec-coframe}, we construct coframe basis 1-forms and ASD Ricci-flat metric which yields an explicit description of anti-self-dual gravity in terms of solutions of the asymmetric heavenly equation.

In section \ref{sec-real}, we impose reality conditions on the equation and the metric and determine the signature of the real metric to be neutral (ultra-hyperbolic) for any solution of the real version of the asymmetric heavenly equation.

In section \ref{sec-cubic}, we find the simplest nontrivial polynomial solution of equation \eref{asym} which is a cubic polynomial that depends on 18 free parameters.

In section \ref{sec-curva}, we locate singularities of our metric and the corresponding Riemann curvature. We are also able to
present explicitly Riemann curvature 2-forms in the particular case of a cubic solution to modified heavenly equation \eref{modheav}. We have
obtained curvature 2-forms in the general case of the asymmetric heavenly equation \eref{asym} but the results are too
lengthy to be presented here. A family of heavenly metrics governed by the asymmetric heavenly equation is explicitly
presented for a particular simple choice of the cubic solution to the latter equation. Riemann curvature 2-forms are also given explicitly in this particular case.

In section \ref{sec-symm}, we present all point symmetries of equation \eref{asym} in the generic case $a\cdot c\ne 0$ and also all such symmetries in the special case $a=c=0$, $b=1$ that corresponds to equation \eref{modheav}. We perform a complete analysis of invariance condition for the cubic solution with respect to all point symmetries of the equation and obtain the result that our solution is invariant under two independent 1-parameter Lie symmetry subgroups. We determine explicitly generators of these subgroups which, although depend on all 18 parameters in our solution, cannot be annihilated by any choice of these parameters. This necessarily implies the existence of two Killing vectors which generate continuous symmetries of our metric \eref{metr} for the solution \eref{cubpoly}.

\section{Lax pair and null tetrad for ASD vacuum metric}
\label{sec-lax}
\setcounter{equation}{0}

Lax pair for asymmetric heavenly equation \eref{asym} has the form
\begin{eqnarray}
& & L_0 = u_{14}\partial_2 - u_{12}\partial_4 + a\partial_3 + \lambda\partial_1 \nonumber
\\ & & M_0 = u_{44}\partial_2 - u_{24}\partial_4 - b\partial_3 - c\partial_1 + \lambda\partial_4 .
\label{lax}
\end{eqnarray}
Indeed, if $\Phi = u_{14}u_{24} - u_{12}u_{44} + au_{34} + bu_{13} + cu_{11}$ is the left-hand side of equation \eref{asym},
we obtain for the commutator of operators $L_0$ and $M_0$
\[ [L_0,M_0] =  \partial_4(\Phi)\partial_2 - \partial_2(\Phi)\partial_4\]
so that $[L_0,M_0] = 0$ on solutions of \eref{asym}. Our Lax pair obviously generalizes the Lax pair of Dubrov and Ferapontov in \cite{fer}.

We use the Lax pair in the Ashtekar-Jacobson-Smolin theorem \cite{ajs} (see also Mason and Newman \cite{mn}). Notation and formulation of results
follow the book of Mason and Woodhouse \cite{MasWood}. Following the ideas in the recent book of Dunajski \cite{dunaj}, we could derive our Lax pair
in a starightforward way, similar to the procedure in \cite{ms}.

 Let $\Omega$ be a holomorphic function of $z^1,z^2,z^3,z^4$.
 Define operators $L$ and $M$ by the relations
\begin{equation}\label{LM}
    L_0 = \Omega L, \quad M_0 = \Omega M
\end{equation}
 with $\Omega$ yet unknown. Then
\begin{equation}\label{commLM}
    [L_0,M_0] = [\Omega L,\Omega M] = 0
\end{equation}
on solutions of \eref{asym}. Define a tetrad $W,Z,\tilde{W},\tilde{Z}$ by splitting operators $L$ and $M$ in the spectral parameter $\lambda$:
 \begin{equation}\label{WZ}
    L = W - \lambda \tilde{Z},\quad M = Z - \lambda \tilde{W}.
 \end{equation}
Let $\nu$ be a holomorphic 4-form on a four-dimensional complex manifold with the coordinates $\{z^i\}$ which
satisfies the following two conditions
\begin{equation}\label{nucond}
    {\mathscr L}_L(\Omega^{-1}\nu) = {\mathscr L}_M(\Omega^{-1}\nu) = 0,
\end{equation}
where ${\mathscr L}$ denotes Lie derivative, and also the normalization condition
\begin{equation}\label{norm}
  24 \nu(W,Z,\tilde{W},\tilde{Z}) = 1,
\end{equation}
Then, according to Proposition 13.4.8 in the book \cite{MasWood}, $W,Z,\tilde{W},\tilde{Z}$ is a null tetrad for the
ASD vacuum metric.

We note that
\[{\mathscr L}_{L_0}(\rmd z^1\wedge \rmd z^2\wedge \rmd z^3\wedge \rmd z^4) =
{\mathscr L}_{M_0}(\rmd z^1\wedge \rmd z^2\wedge \rmd z^3\wedge \rmd z^4) = 0 ,\]
which by virtue of the definition \eref{LM} of $L$ and $M$ is equivalent to
\begin{equation}\label{LMOm}
    {\mathscr L}_{L}(\Omega \rmd z^1\wedge \rmd z^2\wedge \rmd z^3\wedge \rmd z^4) =
  {\mathscr L}_{M}(\Omega \rmd z^1\wedge \rmd z^2\wedge \rmd z^3\wedge \rmd z^4) = 0 .
\end{equation}
Comparing \eref{LMOm} with the condition \eref{nucond}, we conclude that
\begin{equation}\label{nu}
    \nu = \Omega^2 \rmd z^1\wedge \rmd z^2\wedge \rmd z^3\wedge \rmd z^4 .
\end{equation}
From the definition \eref{LM} of $L$ and $M$, we have $L = \Omega^{-1}L_0$ and $M = \Omega^{-1}M_0$
which together with the definition of the Lax pair \eref{lax} and formulae \eref{WZ} yields the explicit
form of the tetrad
\begin{eqnarray}
 & & W = \Omega^{-1}(u_{14}\partial_2 - u_{12}\partial_4 + a\partial_3),\quad \tilde{Z} = - \Omega^{-1}\partial_1 \nonumber
 \\ & & Z = \Omega^{-1}(u_{44}\partial_2 - u_{24}\partial_4 - b\partial_3 - c\partial_1),\quad \tilde{W} = - \Omega^{-1}\partial_4 .
 \label{WZform}
\end{eqnarray}
Using these expressions and the form of $\nu$ in \eref{nu} in the normalization condition \eref{norm}, we obtain $\Omega$:
\begin{equation}\label{OM}
    \Omega = \sqrt{au_{44} + bu_{14}},\quad \mathrm{so\; that}\quad \Delta\stackrel{def}{=}au_{44} + bu_{14} = \Omega^2 > 0.
\end{equation}
The restriction $\Delta=au_{44} + bu_{14}>0$ in \eref{OM} can be easily overcome if we make an odd permutation of the tetrad vectors in
the normalization condition \eref{norm}, e.g. $24 \nu(Z,\tilde{Z},W,\tilde{W}) = 1$.
Then we find that $\Delta = - \Omega^2 <0$. Combining together these two cases, we obtain $ \Omega = \sqrt{|\Delta|}$ and \eref{WZform}
yield the final formulae for the null tetrad
\begin{eqnarray}
 & & W = \frac{1}{\sqrt{|\Delta|}}(u_{14}\partial_2 - u_{12}\partial_4 + a\partial_3),\quad \tilde{Z} = - \frac{1}{\sqrt{|\Delta|}}\,\partial_1\nonumber
 \\ & & Z = \frac{1}{\sqrt{|\Delta|}}(u_{44}\partial_2 - u_{24}\partial_4 - b\partial_3 - c\partial_1),\quad \tilde{W} = - \frac{1}{\sqrt{|\Delta|}}\,\partial_4.
 \label{WZfin}
\end{eqnarray}
According to the result from \cite{MasWood} cited above, formulas \eref{WZfin} define a null tetrad for the
ASD vacuum metric governed by the asymmetric heavenly equation \eref{asym}.

\section{Coframe and ASD metric}
\label{sec-coframe}
\setcounter{equation}{0}

Four basis 1-forms $\omega^i = \omega^i_j \rmd z^j$, which constitute a coframe corresponding to the null tetrad \eref{WZfin},
should satisfy the bi-orthogonality conditions
\begin{equation}\label{ortnorm}
    \omega^1(W) = \omega^2(Z) = \omega^3(\tilde{W}) = \omega^4(\tilde{Z}) = 1
\end{equation}
with all other $\omega^i(W), \omega^i(Z), \omega^i(\tilde{W}), \omega^i(\tilde{Z})$ vanishing.
The coframe 1-forms arise as a solution to these conditions including equations \eref{ortnorm}
\begin{eqnarray}
& &  \omega^1 = \frac{1}{\sqrt{|\Delta}|}\left(b\,\rmd z^2 + u_{44}\,\rmd z^3\right),\quad
  \omega^2 = \frac{1}{\sqrt{|\Delta}|}\left(a\,\rmd z^2 - u_{14}\,\rmd z^3\right) \nonumber \\
& &  \omega^3 = - \frac{1}{\sqrt{|\Delta}|}\left\{(au_{24} + bu_{12})\,\rmd z^2
  + (u_{12}u_{44} - u_{14}u_{24})\,\rmd z^3 + \Delta\, \rmd z^4\right\} \nonumber \\
& &  \omega^4 = - \frac{1}{\sqrt{|\Delta}|}\left(\Delta\,\rmd z^1 + ac\,\rmd z^2
  - cu_{14}\,\rmd z^3\right).
  \label{forms}
\end{eqnarray}
The corresponding ASD vacuum metric has the form
\begin{eqnarray}
& &  ds^2 = 2(\omega^2\omega^4 - \omega^1\omega^3) = \frac{2}{|\Delta|}\left\{\alpha_1 (\rmd z^2)^2 + \alpha_2 (\rmd z^3)^2
  + \alpha_3 \rmd z^2\rmd z^3 \right. \nonumber \\
& &  \left. \mbox{} + \Delta (b\,\rmd z^2\rmd z^4 - a\,\rmd z^1\rmd z^2 + u_{14}\,\rmd z^1\rmd z^3 + u_{44}\,\rmd z^3\rmd z^4 )
  \right\}
\label{metr}
\end{eqnarray}
where we have used the notation
\begin{eqnarray}
& & \Delta=au_{44} + bu_{14},\quad \alpha_1 = b(au_{24} + bu_{12}) - a^2c \nonumber\\
& & \alpha_2 = u_{44}(u_{12}u_{44} - u_{14}u_{24}) - cu_{14}^2
\label{not}
\\ & &  \alpha_3 = b(u_{12}u_{44} - u_{14}u_{24}) + u_{44}(au_{24} + bu_{12}) + 2acu_{14}.
\nonumber
\end{eqnarray}
By using a Reduce program with the package EXCALC (Exterior Calculus of Modern Differential Geometry),
we have checked explicitly that the metric \eref{metr} is indeed Ricci-flat on solutions of the asymmetric heavenly equation
and computed Riemann curvature 2-forms. These results are too lengthy to be presented here. In section \ref{sec-curva}, we present curvature 2-forms for our cubic solution, given in section \ref{sec-cubic}, for modified heavenly equation \eref{modheav} .

\section{Reality condition and signature}
\label{sec-real}
\setcounter{equation}{0}

In order to determine signature of the metric, we have to impose reality conditions on the equation \eref{asym}
and the metric \eref{metr} and then to bring the metric to a diagonal form. The only possible reality condition
for the complex equation \eref{asym} is to consider all the variables $z^i$ and $u$ to be real. Metric \eref{metr}
is a quadratic form in differentials $\rmd z^i$. Applying to it a standard algebraic procedure for the reduction
of a general quadratic form to a canonical one, we obtain a diagonal form of the metric \eref{metr}
\begin{equation}\label{diag}
    \rmd s^2 = \frac{2}{|\Delta|}\left[\frac{l_1^2}{\alpha_1} + \frac{l_2^2}{\beta_1} - \frac{l_3^2}{\beta_2}
    + l_4\,(\rmd z^4)^2 \right]
\end{equation}
where
\begin{equation}\label{not2}
    \beta_1 = \alpha_2 - \frac{\alpha_3^2}{4\alpha_1},\quad \beta_2 = \frac{\Delta^2}{4}\left[\frac{1}{\beta_1}
    \left(u_{14} + \frac{a\alpha_3}{2\alpha_1}\right)^2 + \frac{a^2}{\alpha_1}\right]
\end{equation}
$\alpha_i$ and $\Delta$ are defined in \eref{not} and $l_i$ are defined as follows
\begin{eqnarray}
 & &  l_1 = \frac{1}{2}\left[2\alpha_1\,\rmd z^2 + \alpha_3\,\rmd z^3 + \Delta (b\,\rmd z^4 - a\,\rmd z^1)\right] \nonumber \\
 & &  l_2 = \beta_1\,\rmd z^3 + \frac{\Delta}{4\alpha_1}\,(\alpha_{14}\,\rmd z^1 + \alpha_{44}\,\rmd z^4),\quad
      l_3 = \beta_2\,\rmd z^1 + \frac{\Delta^2\beta}{16\alpha_1^2\beta_1}\,\rmd z^4 \nonumber \\
 & &  l_4 = \frac{\Delta^2}{4}\left[\frac{\Delta^2\beta^2}{64\alpha_1^4\beta_1^2\beta_2} - \frac{\alpha_{44}^2}{4\alpha_1^2\beta_1}
   - \frac{b^2}{\alpha_1}\right].
\label{l_i}
\end{eqnarray}
Here we have used the following shorthand notation
\begin{equation}\label{not3}
    \alpha_{14} = 2u_{14}\alpha_1 + a\alpha_3,\quad \alpha_{44} = 2u_{44}\alpha_1 - b\alpha_3,\quad
    \beta = \alpha_{14}\alpha_{44} - 4ab \alpha_1\beta_1 .
\end{equation}
Of course, the canonical form to which one may reduce a given quadratic form is by no means uniquely determined:
any quadratic form may be diagonalized in many different ways which will affect $l_i$, $\alpha_i$ and $\beta_i$ in \eref{diag}.
But according to the \textit{law of inertia for real quadratic forms}, the number of positive squares
and negative squares in its diagonal form do not depend on the choice of reducing transformation which agrees with the well-known
fact that signature is an invariant.

The form \eref{diag} is still too complicated to determine its signature. However, we can immediately conclude that to
have Euclidean signature, we need $\alpha_1$ and $\beta_1$ to be of the same sign and then $\beta_2$ will have the same sign
too, so the term with $l_3^2$ will have an opposite sign. Therefore, unfortunately we will not have the case of Euclidean signature
$(++++)$.

  For an independent check of our result, we will explicitly show that we indeed have neutral signature. For this purpose, using
relations between $\alpha_i$ and $\beta_i$ we simplify essentially coefficients of the squares in the metric \eref{diag} as
follows
\begin{equation}\label{simple}
    \rmd s^2 = \frac{2}{|\Delta|}\left[\frac{l_1^2}{\alpha_1} - \frac{4\alpha_1 l_2^2}{\Delta^2(u_{24}^2 + 4cu_{12})}
    + \frac{(u_{24}^2 + 4cu_{12})}{u_{12}\Delta^2}\,l_3^2 - \frac{(\rmd z^4)^2}{4u_{12}}\right].
\end{equation}
We now list all possible cases and present the corresponding signatures with the signs following the order of terms in \eref{simple}.
We assume that $\alpha_1>0$ since otherwise we will just get an overall minus sign in the metric.
\begin{enumerate}
 \item $u_{12}>0,\; u_{24}^2 + 4cu_{12} > 0\,:\phantom{\Rightarrow c>0}\quad (+-+\,-)$.
 \item $u_{12}<0,\; u_{24}^2 + 4cu_{12} > 0\,:\phantom{\Rightarrow c<0}\quad (+\,--\,+)$.
 \item $u_{12}>0,\; u_{24}^2 + 4cu_{12} < 0\Rightarrow c<0\,:\quad (++\,-\,-)$.
 \item $u_{12}<0,\; u_{24}^2 + 4cu_{12} < 0\Rightarrow c>0\,:\quad (++++)$.
\end{enumerate}
The first three cases obviously correspond to the neutral signature. The fourth case seems to result in Euclidean signature but
this would contradict our previous conclusion that when the first two terms in \eref{diag} are of the same sign, then the
third term will have an opposite sign. Indeed, studying relations between the quantities that we have introduce, we arrive at
the formula
\[u_{12} = \frac{1}{4\Delta^2\alpha_1}\left(\alpha_{14}^2 + 4a^2\alpha_1\beta_1\right).\]
This shows that when the first two terms in \eref{simple} (and hence in \eref{diag}) have the same sign, which implies that
$\alpha_1>0$ and $\beta_1>0$ simultaneously, $u_{12}$ cannot be negative. Therefore, case (iv) is not realized and we end up
with the neutral signature.

For completeness, we present also the particular case $a=c=0$, $b=1$ which corresponds to the modified heavenly equation in \cite{fer}
\begin{equation}\label{modheav}
    u_{14}u_{24} - u_{12}u_{44} + u_{13} = 0  .
\end{equation}
In this particular case \eref{simple} reduces to the following ASD vacuum metric governed by solutions of \eref{modheav}
\begin{equation}\label{fermetr}
    \rmd s^2 = \frac{2}{u_{12}|u_{14}|^3}\left[u_{14}^2 l_1^2 - \frac{4u_{12}^2}{u_{24}^2}\, l_2^2
    + u_{24}^2 l_3^2 - \frac{u_{14}^2}{4}\, (\rmd z^4)^2\right]
\end{equation}
which for $u_{12}>0$ obviously has the neutral signature $(+-+\,-)$, while $u_{12}<0$ only changes the overall sign.

It is surprising that there is only one real cross section of the asymmetric heavenly equation which leads to the ASD vacuum metric
\eref{simple} with the neutral signature. For the well-known example of the first heavenly equation of Pleba\~nski, there are
two real cross sections: elliptic and hyperbolic versions of the complex Monge-Amp\`ere equation.
We wonder what equation could be the `partner' of the asymmetric heavenly equation such that its real cross section
will govern ASD Ricci-flat metric with Euclidean signature?

\section{Cubic solution of asymmetric heavenly\\ equation}
\label{sec-cubic}
\setcounter{equation}{0}

One of the advantages of describing ASD vacuum gravity by the asymmetric heavenly equation is an easiness of
obtaining polynomial solutions of the latter equation. Here we consider the simplest nontrivial case
(with nonconstant curvature tensor) of a cubic polynomial solution of the general form
\begin{eqnarray}
& &   u = c_1(z^1)^3 + (c_2z^2 + c_3z^3 + c_4z^4)(z^1)^2 \nonumber \\
& &  \mbox{} + \left(c_5(z^2)^2 + c_6(z^3)^2 + c_7(z^4)^2 +c_8z^2z^3 + c_9z^2z^4 + c_{10}z^3z^4\right)z^1 \nonumber \\
& &  \mbox{} + c_{11}(z^1)^2 + c_{12}z^1z^2 + c_{13}z^1z^3 + c_{14}z^1z^4 + c_{15}(z^2)^3 \nonumber \\
& &  \mbox{} + \left(c_{16}z^3 + c_{17}z^4\right)(z^2)^2 + \left(c_{18}(z^3)^2 + c_{19}(z^4)^2 + c_{20}z^3z^4\right)z^2
\label{cubpoly}
\\ & & \mbox{} + c_{21}(z^2)^2 + c_{22}z^2z^3 + c_{23}z^2z^4 + c_{24}(z^3)^3 \nonumber
\\ & & \mbox{} + \left(c_{25}z^3 + c_{26}z^4\right)z^3z^4 + c_{27}(z^3)^2 + c_{28}z^3z^4 + c_{29}(z^4)^3 + c_{30}(z^4)^2  \nonumber.
\end{eqnarray}
The expression \eref{cubpoly} satisfies the asymmetric heavenly equation \eref{asym} when ten coefficients are expressed in terms of other
eighteen ones, which remain as free parameters in the solution, while two other coefficients vanish. With the notation $\delta=3c_{17}c_{29}-c_{19}^2$, the relations between the coefficients of the solution read
\begin{eqnarray}
& &  c_2 = c_{19}\frac{c_5^2}{c_{17}^2},\quad c_3 = \frac{3c}{bc_{17}^3}(c_{29}c_5^3 - c_1c_{17}^3),\quad
     c_4 = 3c_{29}\left(\frac{c^5}{c_{17}}\right)^2 ,\quad c_7 = \frac{3c_5c_{29}}{c_{17}} \nonumber \\
& &  c_9 = \frac{2c_5c_{19}}{c_{17}},\quad c_{10} = 0,\quad c_{20} = \frac{c_{17}c_8}{c_5},\quad c_{26} = 0, \nonumber \\
& &  c_{13} = \frac{1}{b}(2c_{12}c_{30} - c_{14}c_{23} - ac_{28} - 2cc_{11}),\nonumber\\
& &  c_{30} = \frac{c_{17}}{4c_5^2\delta}[3c_8c_{29}(ac_{17} + bc_5) + 2c_5c_{14}\delta]\nonumber \\
& &  c_{23} = \frac{1}{2c_5^2c_{17}\delta}\left[c_8c_{19}c_{17}^2(ac_{17} + bc_5) + 2c_5\delta(c_{12}c_{17}^2 - cc_5^2)\right]\nonumber \\
& &  c_6 = \frac{3c^2}{b^2}\left(c_1 - c_{29}\frac{c_5^3}{c_{17}^3}\right) - \frac{a}{b}c_{25}
              + \frac{3c_8^2c_{29}c_{17}}{4bc_5^2\delta}(ac_{17} + bc_5)
\label{coefsol}
\end{eqnarray}
where $b\cdot c_5\cdot c_{17}\cdot\delta \ne 0$.

Thus, formula \eref{cubpoly} with coefficients determined by \eref{coefsol} yields a solution of asymmetric heavenly equation, in the form of a cubic polynomial  depending on eighteen parameters.

In the particular case of the modifed heavenly equation \eref{modheav}, when $a=c=0$, $b=1$ the coefficients
$c_2, c_4, c_7, c_9, c_{10}, c_{20}, c_{26}$ remain the same, while the other ones simplify as follows
\begin{eqnarray}
 & &  c_3 = 0,\quad c_6 = \frac{3c_8^2c_{29}c_{17}}{4c_5\delta},\quad c_{13} = \frac{c_8c_{17}}{2c_5\delta}(3c_{12}c_{29} - c_{14}c_{19}) \nonumber \\
 & &  c_{23} = \frac{c_{17}}{2c_5\delta}(c_8c_{19} + 2c_{12}\delta) ,\quad c_{30} = \frac{c_{17}}{4c_5\delta}(3c_8c_{29} + 2c_{14}\delta).
 \label{simplcoef}
\end{eqnarray}
Our solution depends on 18 free parameters: $c_1,c_5,c_8,c_{11},c_{12},c_{14},c_{15},c_{16},c_{17}$, $c_{18},c_{19},c_{21},c_{22},c_{24},c_{25},c_{27},c_{28},c_{29}$. In the particular case of equation \eref{modheav}, the number of parameters remains the same.

\section{Family of ASD vacuum metrics and\\ curvature 2-forms}
\label{sec-curva}
\setcounter{equation}{0}

Using solution \eref{cubpoly}, \eref{coefsol} for $u$ in the general form of our metric \eref{metr}, we obtain a family of ASD vacuum metrics depending on eighteen parameters. The expression for this family of metrics is too lengthy to be exhibited even in the particular case $a=c=0$, $b=1$ but we can locate
its singularities, which are zeros of its denominator, as a simple pole at $\Delta=0$. Using our solution for $u$ with generic $a,b,c$, we obtain the explicit equation for
the singular manifold of the metric
\begin{equation}\label{sing}
    2c_5\delta\left\{2c_5\left[c_{17}c_{19}z^2 + 3c_{29}(c_5z^1 + c_{17}z^4)\right] + c_{14}c_{17}^2\right\} + 6ac_8c_{29}c_{17}^3 = 0
\end{equation}
which determines a hyperplane in the four-dimensional space parallel to $z^3$ axis. Another condition for singularities is
$ac_{17} + bc_5 = 0$ which we can always avoid by a suitable choice of $c_{17}$ and $c_5$ for $a$ and $b$ fixed in our equation \eref{asym}.

Now consider a simpler case when $a=c=0$ and $b=1$ of the metric governed by the modifed heavenly equation \eref{modheav}. The equation \eref{sing}
for the singular hyperplane simplifies as follows
\begin{equation}\label{simp}
   D \stackrel{def}{=} 2c_5\left[c_{17}c_{19}z^2 + 3c_{29}(c_5z^1 + c_{17}z^4)\right] + c_{14}c_{17}^2 = 0.
\end{equation}

In this case we are also able to present Riemann curvature 2-forms. We introduce the shorthand notation
\begin{equation}\label{num}
    N = 2c_5\delta z^2 + 3c_8c_{17}c_{29}z^3 + c_{17}(3c_{12}c_{29} - c_{14}c_{19})
\end{equation}
and the 1-forms
\begin{equation}\label{newforms}
    \omega^{12} = c_5\omega^1 + c_{17}\omega^2,\quad \omega^{34} = c_{17}\omega^3 + c_5\omega^4.
\end{equation}
Then the curvature 2-forms read
\begin{eqnarray}
& &   R^2_{\;\;3} = 36 c_5^2 c_{17} c_{29}\frac{N}{{D}^3} \,\omega^{12}\wedge\omega^{34},\quad
      R^1_{\;\;3} = R^2_{\;\;4} = - \frac{c_{17}}{c_5}\,R^2_{\;\;3} \nonumber \\
& &   R^1_{\;\;4} = \frac{c_{17}^2}{c_5^2}\,R^2_{\;\;3},\quad R^1_{\;\;2} = R^3_{\;\;4} = 0.
\label{curv}
\end{eqnarray}
We observe that the pole of the metric at $D = 0$ in \eref{simp} remains in the Riemann curvature tensor but it becomes a multiple pole
of order three.

Further simplifications of the metric are possible by choosing appropriate values of some of the free coefficients
in our solution to the asymmetric heavenly equation. As an example illustrating the form of the metric, let us set to zero all coefficients of the solution except $c_5, c_{17}$ and $c_{19}, c_{29}$ which, by the definition $\delta=3c_{17}c_{29}-c_{19}^2$, satisfy our restriction for the solution:  $c_5\cdot c_{17}\cdot\delta \ne 0$. With the shorthand notation
\begin{eqnarray*}
& & x = 3c_5c_{29}z^1 + c_{17}\left(c_{19}z^2 + 3c_{29}z^4\right),\quad y = c_5c_{19}z^1 + c_{17}\left(c_{17}z^2 + c_{19}z^4\right)\\
& &   \varepsilon = \textrm{sign}(ac_{17} + bc_5)
\end{eqnarray*}
the metric \eref{metr} becomes
\begin{eqnarray}
& &   ds^2 = \frac{\varepsilon}{|x|}\left\{(2by - acc_{17})(\rmd z^2)^2 + \frac{2}{c_{17}}x(2y + cc_5)\rmd z^2\rmd z^3 \right. \nonumber \\
& &   \left. \mbox{} + 2x\left[b\,\rmd z^2\rmd z^4 - a\,\rmd z^1\rmd z^2 + \frac{2}{c_{17}^2} x (c_5\rmd z^1 + c_{17}\rmd z^4)\rmd z^3\right] \right\}.
\label{simpmetr}
\end{eqnarray}
Riemann curvature 2-forms for the metric \eref{simpmetr} read
\begin{eqnarray}\label{curvsimp}
     R^2_{\;\;3} = \frac{9 c_5^2 c_{17} c_{29}(2b\delta z^2 - 3acc_{29})}{2(ac_{17} + bc_5)^2x^3} \,\omega^{12}\wedge\omega^{34},\quad
    R^1_{\;\;3} = R^2_{\;\;4} = - \frac{c_{17}}{c_5}\,R^2_{\;\;3} \\
    R^1_{\;\;4} = \frac{c_{17}^2}{c_5^2}\,R^2_{\;\;3},\quad R^1_{\;\;2} = R^3_{\;\;4} = 0.
\end{eqnarray}
Here again the simple pole of the metric \eref{simpmetr} at $x=0$ converts to a pole of order three at the same location.

We note that the relations between the different curvature 2-forms in \eref{curvsimp} are exactly the same as in \eref{curv} for the modified heavenly equation, whereas the form $R^2_{\;\;3}$ and hence all other nonzero forms are different. It should also be mentioned that the vanishing of the two curvature 2-forms $R^1_{\;\;2} = R^3_{\;\;4} = 0$ holds for all solutions of the asymmetric heavenly equation.

\section{Symmetries of asymmetric heavenly\\ equation}
\label{sec-symm}
\setcounter{equation}{0}

To decide if our solution \eref{cubpoly} with conditions \eref{coefsol} are either invariant or noninvariant solutions, i.e. whether they admit any symmetries of the equation \eref{asym} or not, we need to find all symmetries of the latter equation.

Here we first present the generators of all point symmetries of asymmetric heavenly equation in the generic case when $a\cdot b\cdot c\ne 0$
(from now on, subscripts after commas denote partial derivatives with respect to corresponding $z^i$, primes denote ordinary derivatives)
\begin{eqnarray}
\hspace*{-8mm} & & X_1 = z^2 \partial_2 + u \partial_u,\quad X_2 = - f,_{2}(z^2,z^3)\partial_4 + \zeta f,_{3}(z^2,z^3)\partial_u,
  \quad X_3 = g(z^2,z^3)\partial_u  \nonumber \\
\hspace*{-8mm} & & X_\infty = \left(\frac{1}{2a} H'(z) + \frac{c}{b} m(z^3)\right)\partial_1
  - \left(N,_3(z^2,z^3) + \frac{bd}{ac} z^2\right)\partial_2 \nonumber \\
\hspace*{-8mm} & & \mbox{} + m(z^3)\partial_3 + \frac{1}{2b}\left[H'(z) + \frac{2ac}{b}m(z^3) - 2\left(m'(z^3) + \frac{bd}{ac}\right)\zeta\right]\partial_4
\label{point} 
\\ \hspace*{-8mm} & & \mbox{} + \left[F(z) + \frac{ac}{2b^2}z^2H'(z) + \frac{\zeta^2}{2b}N,_{33}(z^2,z^3) - \frac{a^2c}{b^2}z^1N,_3(z^2,z^3)
  - 2dz^2z^4 \right. \nonumber \\
\hspace*{-8mm} & & \left. \hspace*{15mm} \mbox{} + \frac{bd}{ac} u\right]\partial_u \nonumber
\end{eqnarray}
where $z = bz^1 - cz^3$, $\zeta = az^1 - bz^4$, and $F,H,m,N$ are arbitrary smooth functions, $d$ is an arbitrary constant,
whereas $a,b,c$ are given coefficients of equation \eref{asym}.

In the special case of modified heavenly equation \eref{modheav} when $a=c=0$, $b=1$, the point symmetries become
\begin{eqnarray}
   & &  X_1 = 2z^2\partial_2 - z^4\partial_4,\quad X_2 = z^2\partial_2 + u\partial_u,\quad X_3 = g(z^3)\partial_2 - \frac{(z^4)^2}{2}g'(z^3)\partial_u \nonumber
\\ & &  X_4 = l(z^1)\partial_u,\quad X_5 = m(z^1)\partial_1 \nonumber 
\\ & &  X_6 = - n'(z^3)z^2\partial_2 + n(z^3)\partial_3 + \frac{z^2(z^4)^2}{2} n''(z^3)\partial_u \nonumber
\\ & &  X_7 = h,_2(z^2,z^3)\partial_4 + h,_3(z^2,z^3)z^4\partial_u , \quad X_8 = f(z^2,z^3)\partial_u 
\label{modsym}
\end{eqnarray}
where $g, l, m, n, f, h$ are arbitrary smooth functions.

For simplicity, we choose here the special case of symmetries \eref{modsym} of equation \eref{modheav} with the solution of the same form \eref{cubpoly}
but with simplified expressions for the coefficients \eref{simplcoef}. The condition for invariance of this solution under an arbitrary linear combination
of the symmetry generators \eref{modsym} with arbitrary constant coefficients $C_1, C_2$ has the general form
\begin{equation}\label{inv}
    \left(C_1X_1 + C_2 X_2 + X_3 + X_4 + X_5 + X_6 + X_7 + X_8\right)\left(P(z^i) - u\right)|_{u=P} = 0
\end{equation}
where $P(z^i)$ is our cubic polynomial solution \eref{cubpoly} with some of the coefficients determined by \eref{simplcoef}.
With expressions \eref{modsym} for the symmetry generators, the condition \eref{inv} becomes
\begin{eqnarray}
& &  m(z^1)P,_1 + \left[(2C_1 + C_2)z^2 + g(z^3) - z^2 n'(z^3)\right]P,_2 + n(z^3)P,_3 \nonumber
\\ & &  \mbox{} + \left(h,_2(z^2,z^3) - C_1z^4\right)P,_4 - C_2 P(z^i) + \frac{(z^4)^2}{2}\left[g'(z^3) - z^2n''(z^3)\right] \nonumber
\\ & &  \mbox{} - l(z^1) - z^4 h,_3(z^2,z^3) - f(z^2,z^3) = 0.
\label{invcon}     
\end{eqnarray}
The dependence of \eref{invcon} on $z^4$ is known explicitly, so we can split this equation into several ones by requiring that coefficients of different powers of $z^4$ vanish. Vanishing of the coefficient of $z^1(z^4)^2$  determines $m'(z^1)$. Equating to zero the coefficient of $z^2(z^4)^2$ yields a relation between $h(z^2,z^3)$ and derivatives of $n(z^3)$. Proceeding in a similar way and always preferring restrictions on symmetries to those on coefficients of our solution, we finally end up with a general solution of \eref{invcon} where only two additive constants in $n$ and $m$, that is $n_0$ and $m_0$, are left undetermined whereas all other arbitrary constants and functions in the symmetry generator in \eref{inv} are uniquely determined. This means that generically, i.e. with no further restrictions on the coefficients of the solution, there are exactly two independent symmetry generators which leave our cubic solution invariant and $X$ that satisfies \eref{invcon} has the form $X = n_0 K_1 + m_0 K_2$, where $K_1$ and $K_2$ are the two independent symmetry generators that leave our solution invariant. This fact implies that there are two Killing vectors which generate continuous symmetries of the ASD metric that is governed by the cubic solution.

Here we present explicit forms of symmetries $K_1$ and $K_2$.
\begin{eqnarray}
   & & K_1 = \partial_3 + \frac{c_8c_{17}}{2c_5\delta} (c_{19}\partial_4 - 3c_{29}\partial_2) + \frac{1}{2c_5^2\delta}\left[z^4h_1,_3(z^2,z^3)\right.
\label{k1}  
\\ & & \left.\mbox{} + \frac{1}{2c_{17}}\left(K^{(1)}_{22}(z^2)^2 + K^{(1)}_{33}(z^3)^2 + 2K^{(1)}_{23}z^2z^3 + K^{(1)}_{2}z^2 + K^{(1)}_{3}z^3\right)\right]\partial_u , \nonumber
\end{eqnarray}
\begin{eqnarray}
& & K_2 = \partial_1 - \frac{c_5}{c_{17}}\partial_4 + \left\{\frac{1}{c_{17}^3}\left[3(c_1c_{17}^3 - c_{29}c_5^3)(z^1)^2
    + c_{17}^2(2c_{11}c_{17} - c_5c_{14})z^1\right]\right. 
 \nonumber       
\\ & & \left. \mbox{} - \frac{3c_8c_{29}}{2\delta}z^4
    + \frac{1}{4c_5c_{17}\delta}\left(K^{(2)}_{33}(z^3)^2) + K^{(2)}_{2}z^2 + K^{(2)}_{3}z^3\right)\right\}\partial_u
\label{k2}
\end{eqnarray}
with the following notation for the constants
\begin{eqnarray}
& &   K^{(1)}_{22} = 2c_5c_{17}[2c_5c_{16}\delta + c_8c_{17}(c_{17}c_{19} - 9c_{15}c_{29})] \nonumber \\
& &   K^{(1)}_{33} = c_5\left[12c_5c_{17}c_{24}\delta + 2c_8c_{17}^2(c_{19}c_{25} - 3c_{18}c_{29})\right] \nonumber \\
& &   K^{(1)}_{23} = c_{17}\left[4c_5^2c_{18}\delta + c_8c_{17}(c_8c_{17}c_{19} - 6c_5c_{16}c_{29}\right] \nonumber \\
& &   K^{(1)}_{2} = c_{17}\left[4c_5^2c_{22}\delta + 2c_8c_{17}(c_{12}c_{17}c_{19} - 6c_5c_{21}c_{29}) + \frac{1}{\delta}c_8^2c_{17}^2c_{19}^2\right] \nonumber \\
& &   K^{(1)}_{3} = 2c_5c_{17}[4c_5c_{27}\delta - c_8c_{17}(3c_{22}c_{29} - c_{19}c_{28})] \nonumber \\
& &   K^{(2)}_{33} = 3c_8^2c_{17}^2c_{29} - 4c_5^2c_{25}\delta ,\quad K^{(2)}_{2} = - 2c_5c_8c_{17}c_{19} \nonumber \\
& &   K^{(2)}_3    = 2c_8c_{17}^2\gamma - 4c_5^2c_{28}\delta
\label{Kij}
\end{eqnarray}
where $\delta = 3c_{17}c_{29} - c_{19}^2$,\quad $\gamma = 3c_{12}c_{29} - c_{14}c_{19}$. Coefficients \eref{Kij} determine function $f(z^2,z^3)$ in generator $X_8$ in \eref{modsym}. Function $h(z^2,z^3)$ in generator $X_7$ in \eref{modsym} has the form $h = n_0h_1(z^2,z^3) + m_0h_2(z^2,z^3) + h_0$, where
\begin{eqnarray}
& & h_1 = \frac{1}{4c_5^2\delta}\left[(4c_5^2c_{25}\delta - 3c_8^2c_{17}^2c_{29})(z^3)^2
    + \frac{2}{c_{17}}(2c_5^2c_{28}\delta - c_8c_{17}^2\gamma)z^3 \right.  \nonumber \\
& & \left. \mbox{} + 2c_5c_8c_{17}c_{19}z^2\right],\quad h_2 = - \frac{1}{2c_{17}\delta}(2c_5\delta z^2 + 3c_8c_{17}c_{29}z^3)
\label{h1}    
\end{eqnarray}
and $h_0$ is an arbitrary constant not involved in the symmetry generators $K_1$ and $K_2$ containing only derivatives of $h$. We note that generators
$K_1$ and $K_2$ in \eref{k1} and \eref{k2}, respectively, depend on the coefficients of our solution \eref{cubpoly}. This means that changing the solution by varying the coefficients, we also change the symmetry generators which leave the solution invariant.

For the simple particular case of the solution that governs metric \eref{simpmetr}, symmetries of the solution simplify as follows
\[K_1 = \partial_3 ,\quad K_2 = \partial_1 - \frac{c_5}{c_{17}}\,\partial_4 - 3c_{29}\frac{c_5^3}{c_{17}^3}\left(z^1\right)^2\partial_u  \]
and correspond again to two Killing vectors of the metric \eref{simpmetr}, though this solution may admit more symmetries and more Killing vectors for this metric.

\section*{Conclusion}

We have obtained a description of anti-self-dual gravity in terms of solutions of the asymmetric heavenly equation. Since so far this was the only one out of inequivalent heavenly equations for which such a description was missing, we thus have filled in the gap in describing ASD vacuum gravity metrics in terms of scalar PDEs for a potential of the metric, following the original idea of Pleba\~nski \cite{pleb}. We were able to do this by using the Ashtekar-Jacobson-Smolin-Mason-Newman theorem \cite{ajs,mn,MasWood} as our basic tool. We have obtained a null tetrad, coframe and ASD Ricci-flat metric defined on solutions of the asymmetric heavenly equation.

By imposing a reality condition, we were able to determine the signature of the real cross-section of the metric which turned out to be a neutral signature for any solution of the equation. This is a surprising fact because for all other heavenly equations there always exist two real cross-sections, one corresponding to Euclidean signature and the other one to neutral signature of the corresponding ASD vacuum metric. This raises a question: how to find a natural `partner' for the asymmetric heavenly equation which, after imposing reality conditions, would determine a real ASD vacuum metric with Euclidean signature?

We found that the asymmetric heavenly equation was especially convenient for finding polynomial solutions. As a particular illustration of this property, we have obtained the simplest nontrivial polynomial solution, which is a general form of a cubic polynomial in all four variables that satisfies the equation and depends on 18 free parameters. Our results are essentially simplified in the particular case of the modified heavenly equation of Dubrov and Ferapontov \cite{fer}. In this case, we were able not only locate singularities of the metric and Riemann curvature, which are simple and third-order poles respectively on a hyperplane parallel to $z^3$ axis, but also present explicit expressions for the curvature 2-forms. Explicit expressions for the metric and curvature 2-forms were also given in the particular case of the cubic solution to the asymmetric heavenly equation, where we have annihilated all parameters with the exception of four of them.

In order to check whether our solution is invariant or noninvariant one, we have determined all point symmetries of the asymmetric heavenly equation and analyzed the condition for our solution to be invariant under an arbitrary one-parameter subgroup of the Lie symmetry group. We have found that there are two independent symmetries of the equation which leave the cubic solution invariant and presented an explicit form of the symmetry generators. This implies the existence of two Killing vectors which generate continuous symmetries of our metric.

The work on higher-order polynomial solutions, which hopefully will admit less Killing vectors if any, is currently in progress.  We will also concentrate now on a search for an appropriate `partner' of the asymmetric heavenly equation with a real version that determines ASD vacuum metric with Euclidean signature. For such an equation, we plan to construct noninvariant solutions by using partner symmetries, the existence of which is a specific property of heavenly equations \cite{shma}, in order to obtain ASD vacuum metrics of Euclidean signature with no Killing vectors as a step forward for finding the metric of the gravitational $K3$ instanton \cite{dunaj}.

\section*{Acknowledgements}

The research of MBS was supported in part by the research grant
from Bo\u{g}azi\c{c}i University Scientific Research Fund (BAP), research
project No. 6324.


\begin{thebibliography}{99}
\bibitem{pleb}
  Pleba\~nski J F 1975  \emph{J. Math. Phys.} \textbf{16} 2395--402
\bibitem{husain}
  Husain V 1994 \emph{Class. Quantum Grav.} \textbf{11} 927--937 \emph{arXiv:gr-qc/9310003}
\bibitem{shma}
  Sheftel M B and Malykh A A 2009 \emph{J. Phys. A: Math.~Theor.} \textbf{42} 395202
\bibitem{fer}
  Doubrov B and Ferapontov E V 2010 \emph{J. Geom. Phys.} \textbf{60} 1604--1616; \emph{arXiv:0910.3407v2}
\bibitem{ms}
  Malykh A A and Sheftel M B 2011 \emph{J.~Phys.~A: Math.~Theor.} \textbf{44} 155201; \emph{arXiv:1011.2479v3}
\bibitem{dy}
  Yaz{\i}c{\i} D 2011 \emph{J. Phys. A: Math.~Theor.} \textbf{44} 505203
\bibitem{ajs}
Ashtekar A, Jacobson T and Smolin L 1988 \emph{Commun. Math. Phys.} \textbf{115} 631--648
\bibitem{mn}
Mason L J and Newman E T 1989 \emph{Commun. Math. Phys.} \textbf{121} 659--668
\bibitem{MasWood}
Mason L J and Woodhouse N M J 1996 \emph{Integrability, Self-Duality, and Twistor Theory},\\
London Mathematical Society Monographs New Series (Oxford: Clarendon Press)
\bibitem{dunaj}
Dunajski M 2010 \emph{Solitons, Instantons and Twistors} (Oxford: University Press)
\end{thebibliography}
\end{document}